\documentclass[a4paper, preprint, pre]{revtex4-1}
\usepackage[english]{babel}
\usepackage[utf8]{inputenc}
\usepackage{verbatim}
\usepackage{graphicx}
\usepackage{graphics}
\usepackage{color}
\usepackage{textcomp}
\usepackage{amsfonts}
\usepackage{amsmath}
\usepackage{amssymb}
\usepackage{mathtools}
\usepackage{tabularx}
\newcolumntype{Y}{>{\centering\arraybackslash}X}

\begin{document}

\title{Epidemic Outbreaks on Quenched Scale-Free Networks}

\author{D. S. M. Alencar}
\affiliation{Departamento de F\'{\i}sica, Universidade Federal do Piau\'{i}, 57072-970, Teresina - PI, Brazil}
\author{T. F. A. Alves}
\affiliation{Departamento de F\'{\i}sica, Universidade Federal do Piau\'{i}, 57072-970, Teresina - PI, Brazil}
\author{F. W. S. Lima}
\affiliation{Departamento de F\'{\i}sica, Universidade Federal do Piau\'{i}, 57072-970, Teresina - PI, Brazil}
\author{G. A. Alves}
\affiliation{Departamento de F\'{i}sica, Universidade Estadual do Piau\'{i}, 64002-150, Teresina - PI, Brazil}
\author{A. Macedo-Filho}
\affiliation{Departamento de F\'{i}sica, Universidade Estadual do Piau\'{i}, 64002-150, Teresina - PI, Brazil}
\author{R. S. Ferreira}
\affiliation{Departamento de Ci\^{e}ncias Exatas e Aplicadas, Universidade Federal de Ouro Preto, 35931-008, Jo\~{a}o Monlevade - MG, Brazil}

\date{Received: date / Revised version: date}

\begin{abstract}

We present a finite-size scaling theory of a contact process with permanent immunity on uncorrelated scale-free networks. We model an epidemic outbreak by an analogue of susceptible-infected-removed model where an infected individual attacks only one susceptible in a time unit, in a way we can expect a non-vanishing critical threshold at scale-free networks. As we already know, the susceptible-infected-removed model can be mapped in a bond percolation process, allowing us to compare the critical behavior of site and bond universality classes on networks. We used the external field finite-scale theory, where the dependence on the finite size enters the external field defined as the initial number of infected individuals. We can impose the scale of the external field as $N^{-1}$. The system presents an epidemic-endemic phase transition where the critical behavior obeys the mean-field universality class, as we show theoretically and by simulations.

\end{abstract}

\keywords{Uncorrelated Configuration Model; Epidemic processes; Dynamic Percolation; Continuous phase transition; Logarithmic corrections; External field.}
\begin{center}

\end{center}
\pacs{}

\maketitle

\section{Introduction}

We consider an analogue of susceptible-infected-susceptible (SIR) model\cite{Cohen-2010, Barabasi-2016}, already defined in lattices where it is called as asynchronous SIR model. The SIR model is a fundamental epidemic spreading model\cite{Keeling-2007} which presents a non-equilibrium phase transition, and describes an epidemic propagation where the time-scale of the disease duration is small compared with the individuals lifetime. In the SIR dynamics, individuals when recover, gain permanent immunity and effectively are excluded from the dynamics, which can be studied with various techniques and approaches: 1) deterministic differential equations in a well stirred population\cite{Keeling-2008}, 2) a birth-death process by means of stochastic differential equations\cite{vanKampen-1981}, and 3) a cellular automaton which is a Markovian process\cite{vanKampen-1981} in a lattice or network.

In particular, one can define the basic reproduction number $\lambda=\mu_c/\mu_r$. The epidemics' outcome in a well-stirring population depends on the value of $\lambda$, which measures the number of secondary contaminations caused by a primary infected individual. If $\lambda < 1$, epidemics vanish exponentially fast and for $\lambda>1$, epidemics grow exponentially and reach a maximal value, then decreases. One can do two strategies to maintain $\lambda<1$. The first is by increasing $\mu_r$, by treatments that improve the individuals' capacity to heal, or by quarantines. The second is decreasing $\mu_c$ by vaccination when disposable or by social distancing.

In this way, the problem of epidemic transition is formally equivalent to a critical threshold in a non-equilibrium continuous phase transition. We introduce the asynchronous SIR model as a contact process with permanent immunity, where a non-zero critical point is an outcome due any infected individual can contaminate only one susceptible neighbor in a time step, unlike its standard definition\cite{Satorras-2001, Moreno-2002}, for which an infected individual can attack any other in its neighborhood which replicates airborne diseases and spreading computer viruses.

In this work, we consider the same stochastic asynchronous update version in uncorrelated power-law networks, where we select one infected individual in a markovian step. As already known, the asynchronous SIR model is on the dynamic percolation universality class\cite{Tome-2010, deSouza-2011, Tome-2011, Tome-2015, Santos-2020, Alencar-2020}. The lattice SIR model is a special case of the susceptible-infected-removed-susceptible (SIRS) model, and also the predator-prey model. We coupled the Asynchronous SIR model\cite{Tome-2010, deSouza-2011, Tome-2011, Tome-2015, Santos-2020, Alencar-2020} to the Barabasi-Albert (BA)\cite{Cohen-2010, Barabasi-2016} networks and the uncorrelated model (UCM)\cite{Catanzaro-2005}. Both BA and UCM networks provide a test of heterogeneous mean-field theory where we can compare simulation and theoretical results.

Interactions are a crucial aspect of the epidemic spreading\cite{Ferguson-2007}, and special features of scale-free networks are superdiffusion and the presence of hubs, describing superspreaders\cite{Barabasi-2016}. The standard SIR model of Refs. \cite{Satorras-2001, Moreno-2002, Boguna-2013} characterizes airborne diseases and computer viruses in scale-free networks, which have a vanishing threshold in the infinite BA network limit\cite{Satorras-2001, Moreno-2002}. However, the prevalence is exponentially small\cite{Satorras-2001, Moreno-2002}, irrespective of the permanent or temporary immunity.

We present results for finite-size scaling corrections to scaling of the asynchronous SIR model, which we expect to be exact on annealed networks. However, we show by simulations, that the theory is also valid for quenched networks as, for example, the BA and UCM networks. A power-law network, besides the degree distribution, is also characterized by the degree correlations. In scale-free networks where degree fluctuations increase with the network size, the degrees of the two connected nodes are not generally independent. Degree-degree correlations are measured by the conditional probability $P\left(k' \mid k\right)$. In addition, also important is the average degree of nearest neighbors
\begin{equation}
  k_\text{nn} = \sum_{k'} k' P\left(k' \mid k\right),
\end{equation}
which is constant for a neutral network. An increasing $k_\text{nn}$ with $k$ defines an assortative network, and the converse for an disassortative network. A scale-free network with unrestricted bonds is naturally disassortative, connecting hubs with poorly connected nodes.

We outline the paper as follows. In Sec.\ \ref{sec:theory}, we describe the finite-size theory. In Sec.\ \ref{sec:results} how we implemented the model and the relevant parameters obtained from the cluster distribution. Finally, we present our conclusions in Sec.\ \ref{sec:conclusions}.

\section{External field finite-size scaling Theory\label{sec:theory}}

In a first step, we develop a finite-size theory for the complete graph, or equivalently, a well-stirring population of $N$ individuals where the finite size enters the external field. For the asynchronous SIR model, and equivalently, the contact process with permanent immunity, the external field is the initial infected population $I(0)$. 

\subsection{Well-stirring graph}

We begin from the mass-action laws for the SIR model, namely, the Kermack and Mc Kendrick system of differential equations\cite{Keeling-2008}
\begin{eqnarray}
  \frac{\partial S}{\partial t} &=& -\mu_c S(t)I(t) \nonumber \\
  \frac{\partial I}{\partial t} &=& -\mu_c S(t)I(t) - \mu_r I(t) \nonumber \\
  \frac{\partial R}{\partial t} &=& \mu_r I(t)
  \label{sir-wellstirring}
\end{eqnarray}
where $\mu_c$ is the contamination rate, $\mu_r$ is the recovery rate, and the susceptible $S(t)$, infected $I(t)$, and removed $R(t)$ compartment densities are related by 
\begin{equation}
  S(t) + I(t) + R(t) = 1.
  \label{constraint}
\end{equation}

We can obtain an approximate stationary solution for the removed compartment by exploiting the fact that the system always evolve to the absorbing state. We can write for the stationary infected compartment the following asymptotic behavior
\begin{equation}
  \lim_{t \to \infty} I(t) = I_\infty = 0,
  \label{absorbing-state}
\end{equation}
We can integrate $R(t)$ from Eq.\ \ref{sir-wellstirring} to write
\begin{equation}
  R(t) = \mu_r \phi(t),
  \label{removed-time-wellstirring}
\end{equation}
and also by integrating $S(t)$ from Eq.\ \ref{sir-wellstirring}, we can write
\begin{equation}
  S(t) = \left( 1-h \right) \exp{\left[-\mu_c\phi(t)\right]},
  \label{susceptible-time-wellstirring}
\end{equation}
where we used the following initial conditions
\begin{equation}
  S(0) = 1-h, \quad I(0) = h, \quad \text{and} \quad R(0) = 0,
  \label{initial-conditions-wellstirring}
\end{equation}
and the following definition
\begin{equation}
   \phi(t) = \int_{0}^{t}I(t')dt'.
   \label{auxiliar-function-wellstirring}
\end{equation}
Now, from the constraint in Eq.\ {\ref{constraint}}, we can write
\begin{equation}
  R_\infty = 1 - S_\infty
\end{equation}
where $R_\infty$, and $S_\infty$ are the asymptotic stationary values of $R(t)$ and $S(t)$, respectively, defined in an analogous way of Eq.\ \ref{absorbing-state}. Finally, by eliminating the auxiliary function $\phi$, we have, from Eqs. \ref{removed-time-wellstirring}, and \ref{susceptible-time-wellstirring}
\begin{equation}
  R_\infty = 1 - \left( 1-h \right) \exp{\left( - \lambda R_\infty \right)},
  \label{removed-wellstirring}
\end{equation}
where $\lambda$ is the basic reproduction number, defined as
\begin{equation}
  \lambda = \frac{\mu_c}{\mu_r}.
  \label{R0}
\end{equation}

We can expand the exponential in the stationary solution, written in Eq.\ \ref{removed-wellstirring}, to obtain up to second order in $R_\infty$
\begin{equation}
  h \approx \left(\lambda_c-\lambda\right) R_\infty + \frac{1}{2} \lambda^2 R_\infty^2,
  \label{stationary-scaling-wellstirring}
\end{equation}
which approximately gives the removed compartment close to $\lambda_c=1$ in a well-stirring population. The system displays an epidemic phase transition at $\lambda=\lambda_c$ with a external field exponent
\begin{equation}
  \delta = 2.
\end{equation}

We can take the route to the thermodynamic limit by imposing the finite-size external field scaling
\begin{equation}
h \propto N^{-1},
\label{field-scaling}
\end{equation}
which means a initial condition with a patient zero in a well-stirring population of susceptibles. Taking $\lambda \approx \lambda_c = 1$, we obtain from Eqs. \ref{stationary-scaling-wellstirring}, and \ref{field-scaling}
\begin{equation}
  R_\infty \approx \left(\lambda^2 N/2\right)^{-1/2},
  \label{removed-scaling-wellstirring}
\end{equation}
and by taking $h=0$ in Eq.\ \ref{stationary-scaling-wellstirring}, combined with Eq.\ \ref{removed-scaling-wellstirring}, we also obtain
\begin{equation}
  \lambda - \lambda_c \approx \left( 2N/\lambda^2 \right)^{-1/2}.
  \label{shifting-scaling-wellstirring}
\end{equation}
We can combine Eqs. \ref{removed-scaling-wellstirring}, and \ref{shifting-scaling-wellstirring} in the following finite-size scaling relation
\begin{equation}
 R_\infty \approx \left(\lambda^2 N/2\right)^{-1/2}F\left[\left(2N/\lambda^2 \right)^{1/2}\left(\lambda - \lambda_c\right) \right].
\label{wellstirring-finitesize-scaling}
\end{equation}

We now couple the asynchronous SIR model to a well-stirring population. However, we do not have a well-defined boundary spanning condition that we can use to identify a percolating cluster. The lack of a proper boundary condition prevents us from determining the order parameter
\begin{equation}
P = \left< P_\infty \right>,
\label{oldorderparameter}
\end{equation}
for the SIR model, defined as the average ratio of the dynamics realizations that leads to a percolating cluster. This also affects the cumulant ratio defined in Ref.\ \cite{deSouza-2011} which depends directly on the order parameter. We should build a new definition of a cumulant that enables us to determine the critical threshold from the cluster distribution moments.

The cluster distribution moments $s^\ell$ of the cluster distribution $n_{\mathrm{cluster}}(s,\lambda)$ 
\begin{equation}
s^\ell = \frac{1}{N_r}\sum_s s^{\ell+1} n_{\mathrm{cluster}}(s,\lambda),
\label{wellstirring-moment-distribution}
\end{equation}
as functions of the final removed population are simply $s^\ell = N_r^{\ell}$ where $N_r$ is the number of removed individuals, because every dynamics realization generates only one cluster with size $s=N_r$. The averages on the dynamics realizations are just $\left< s^\ell \right> = \left< N_r^\ell \right>$, which obey the following finite-size scaling relation\cite{Stauffer-1992, Christensen-2005}
\begin{equation}
  \left< N_r^\ell \right> \approx N^{(\ell-1)\beta/\nu + \ell\gamma/\nu} f_{n}\left[ N^{1/\nu} \left( \lambda - \lambda_c \right) \right],
  \label{wellstirring-moments}
\end{equation}
and from Eq. \ref{wellstirring-moments}, we can see that the cumulant ratio
\begin{equation}
  U = \frac{\left<N_r^2\right>^2}{\left<N_r\right>\left<N_r^3\right>},
  \label{bindercumulant}
\end{equation}
in terms of the cluster moments are universal at the critical threshold. In addition, the removed concentration, defined as
\begin{equation}
  n_r = \left<N_r\right>/N
\end{equation}
should scale as $N^{\gamma'/\nu-1}$. We can also define the following moment ratio
\begin{equation}
  \psi_i = \frac{\left<N_r^2\right>}{\left<N_r\right>^2},
  \label{susceptibility}
\end{equation}
which scales as $N^{\beta/\nu}$. The averages $U$, $n_r$, and $\psi_r$ should obey the following FSS relations\cite{Stauffer-1992, Christensen-2005, deSouza-2011, Santos-2020, Alencar-2020}
\begin{eqnarray}
U      &\approx& f_U\left[ N^{1/\nu} \left( \lambda - \lambda_c \right) \right], \nonumber \\
n_r    &\approx& N^{\gamma'/\nu-1}f_{n_r}\left[ N^{1/\nu} \left( \lambda - \lambda_c \right) \right], \nonumber \\
\psi_r &\approx& N^{\beta/\nu}f_{\psi_r}\left[ N^{1/\nu} \left( \lambda - \lambda_c \right) \right].
\label{fss-relations}
\end{eqnarray}
where we should sum in all generated clusters $s$ by the dynamics. In addition, the expected value of $n_r$ should be $R_\infty$, in a way that Eq.\ \ref{wellstirring-finitesize-scaling} yields the critical mean-field exponents
\begin{equation}
  \nu=2, \quad \text{and} \quad \gamma'=\beta=1.
\end{equation}
when comparing with Eq. \ref{fss-relations}.

To simulate the model in a well-stirring population, we use the SSA algorithm. The susceptible, infected and removed populations are $N_s$, $N_i$, and $N_r$, respectively. We have two reaction channels: the first one has a propensity $p_1=\mu_c N_i N_s/N$, and the other has a propensity $p_2=\mu_r N_i$. We begin the compartments with $N_s=N-1$, $N_i=1$, and $N_r=0$. Then, we update the populations according to the rules:
\begin{enumerate}
  \item [(1)] We calculate the time lapse 
\begin{equation}
\tau = - \ln (1-r)/(p_1+p_2),
\end{equation}
of the reaction at time $t$ by generating a uniform random number $r$ in the interval $[0,1)$. In addition, we update $t$ with $t+\tau$;
  \item [(2)] We select the channel with a probability proportional to its propensity. If we choose the first channel, we update the populations according to 
\begin{equation}
N_s(t+\tau)=N_s(t)-1,  \quad \text{and} \quad N_i(t+\tau)=N_i(t)+1.
\end{equation}
Instead, for the second channel, we update the populations according to 
\begin{equation}
N_i(t+\tau)=N_i(t)-1, \quad \text{and} \quad N_r(t+\tau)=N_r(t)+1.
\end{equation}
  \item [(3)] We update the system until $N_i=0$.
\end{enumerate}

\begin{figure}[p]
\begin{center}
\includegraphics[scale=0.15]{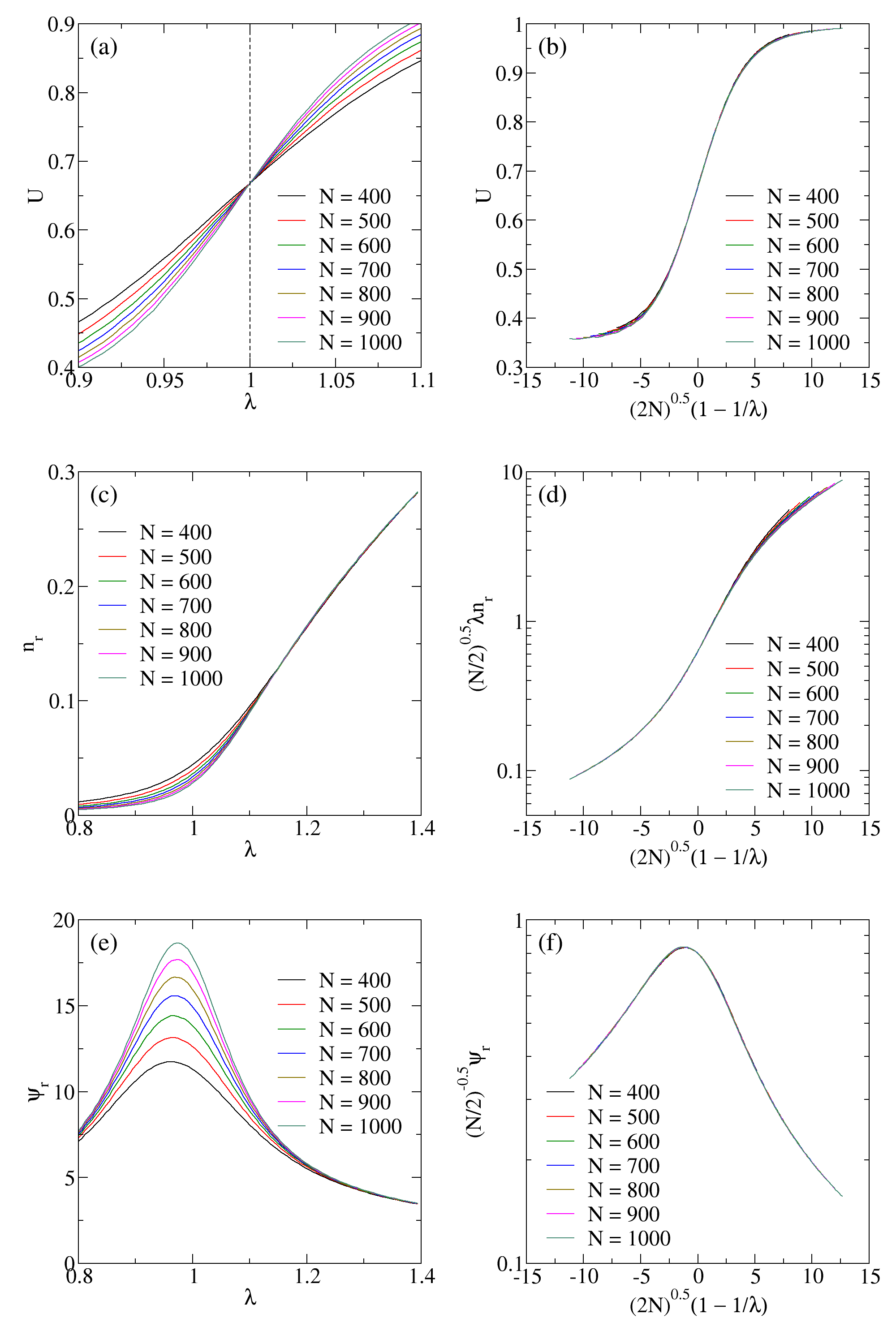}
\end{center}
\caption{(Color Online) We show the stationary simulation results of the fraction of removed individuals $n_r$, the average moment ratio $\psi_r$ defined in Eq.\ \ref{susceptibility}, and the cumulant ratio in Eq.\ \ref{bindercumulant}, in (a), (c), and (e). respectively, for the well-stirring SIR model as functions of the basic reproduction number $\lambda$. We considered a fixed total population $N$ in each curve. We show the data collapses of $n_r$, $\psi_r$, and $U$ in (b), (d), and (f), respectively. The critical threshold for the basic reproduction number in the well-stirring regime reproduces the exact result $\lambda_c = 1$. Data collapses according to Eq.\ \ref{wellstirring-finitesize-scaling} are compatible with $\beta=1$, $\gamma=1$, and shifting exponent $\nu=2$.}
\label{sir-wellstirring-critical}
\end{figure}

We show the results of the SIR model in a well-stirring regime (complete graph) in Fig.\ref{sir-wellstirring-critical}. In (a), we show the ratio in Eq.\ \ref{bindercumulant}, which is universal at the basic reproduction number $\lambda_c=1$. In (c), we show the fraction of removed population $n_r$ where we can see a phase transition between an endemic phase for $\lambda \leq \lambda_c$ and the epidemic phase for $\lambda \ge \lambda_c$, also present in the moment ratio $\psi_r$ in (e). From the cumulant ratio, we could determine the critical threshold that reproduces the exact result for the well-stirring SIR model. Also, we could reproduce the exact critical exponents by data collapses using Eqs. \ref{fss-relations}, which have the values $1/\nu=1/2$, $1-\gamma'/\nu=1/2$, and $\beta/\nu=1/2$, yielding the critical exponents $\beta=1$, $\gamma'=1$, and the shifting exponent $\nu=2$.

\subsection{Uncorrelated power-law network}

The analogous model defined by Eq.\ \ref{sir-wellstirring} coupled to a graph with $N$ nodes, and an adjacency matrix $A_{i,j}$, is defined by the following system of equations, valid for $i=1,2,\cdot,N$ where the mass-action laws for each node are
\begin{eqnarray}
  \dfrac{\partial S_{i}}{\partial t} &=& -\mu_{c} S_{i}(t)\sum_{j}\frac{A_{i,j}}{k_i}I_{j}(t), \nonumber \\
  \dfrac{\partial I_{i}}{\partial t} &=&  \mu_{c} S_{i}(t)\sum_{j}\frac{A_{i,j}}{k_i}I_{j}(t)-\mu_{r} I_{i}(t), \nonumber \\
  \dfrac{\partial R_{i}}{\partial t} &=&  \mu_{c} I_{i}(t),
  \label{sir-generalgraph}
\end{eqnarray}
where $k_i$ is the degree of the node $i$, and the constraint for every node is given by
\begin{equation}
  S_{i}(t) + I_{i}(t) + R_{i}(t) = 1.
  \label{constraint-generalgraph}
\end{equation}
The system of Eqs.\ \ref{sir-generalgraph} is nonlinear and does not have a closed solution. However, we can obtain a stationary solution for a annealed graph. The stationary state is always absorbing and in the regular and random lattices, they are compatible with a geometric phase transition. The geometric phase transition is characterized by the formation of a giant component in the stationary state controlled by the basic reproduction number, given in Eq.\ \ref{R0}.

We consider the system of Eqs.\ \ref{sir-generalgraph} in the particular case of annealed networks, where a heterogeneous mean-field theory is exact, and any node dependent stochastic variable should be dependent only on the node degree. Rewriting the system of Eqs.\ \ref{sir-generalgraph} in the heterogeneous mean-field regime, we have for an particular node
\begin{eqnarray}
  \dfrac{\partial S_{k}}{\partial t} &=& -\mu_{c} S_{k}(t)\sum_{k'}\frac{P_{i,j}(k,k')}{k}I_{k}(t),                    \nonumber \\
  \dfrac{\partial I_{k}}{\partial t} &=&  \mu_{c} S_{k}(t)\sum_{k'}\frac{P_{i,j}(k,k')}{k}I_{k}(t) - \mu_{r} I_{k}(t), \nonumber \\
  \dfrac{\partial R_{k}}{\partial t} &=&                                                             \mu_{r} I_{k}(t).
  \label{sir-generalgraph-hmf}
\end{eqnarray}
However, for statistical independent nodes, we can write
\begin{equation}
  P_{i,j}(k,k') = kP\left( k \mid k' \right)\frac{k'-1}{k'},
  \label{bond-probability}
\end{equation}
which gives the connection probability. In addition, we consider a local tree approximation where a infected node can not reinfect the source of infection, expressed by the factor $(k'-1)/k$ in Eq.\ \ref{bond-probability}\cite{Barrat-2008}. The conditional probability for a neutral, uncorrelated graph is given by
\begin{equation}
  P\left( k \mid k' \right) = \frac{k'P(k')}{\left< k \right>},
  \label{uncorrelated-network}
\end{equation}
where $P(k)$ is the network degree distribution\cite{Barrat-2008, Cohen-2010, Barabasi-2016}. We can combine Eqs. \ref{sir-generalgraph-hmf}, \ref{bond-probability} and \ref{uncorrelated-network} to obtain mass-action laws, exact for uncorrelated annealed networks
\begin{eqnarray}
  \dfrac{\partial S_{k}}{\partial t} &=& -\mu_{c} S_{k} \Theta(t),                    \nonumber \\
  \dfrac{\partial I_{k}}{\partial t} &=&  \mu_{c} S_{k} \Theta(t) - \mu_{r} I_{k}(t), \nonumber \\
  \dfrac{\partial R_{k}}{\partial t} &=&                            \mu_{r} I_{k}(t),
  \label{sir-generalgraph-massaction}
\end{eqnarray}
where
\begin{equation}
  \Theta(t) = \frac{1}{\left< k \right>} \sum_{k} \left( \frac{k-1}{k} \right) P(k)I_k(t).
  \label{auxiliary-function}
\end{equation}

We can find an evolution equation from Eqs.\ \ref{sir-generalgraph-massaction}, starting from the initial conditions
\begin{equation}
  S_k(0) = 1-h, \quad I_k(0) = h, \quad \text{and} \quad R_k(0) = 0,
  \label{initial-conditions-hmf}
\end{equation}
by integrating Eq.\ \ref{sir-generalgraph-massaction} and defining the following auxiliary function, similarly to the well-stirring case
\begin{equation}
  \phi(t) = \int_{0}^{t} \Theta(t')dt'
  \label{auxiliary-function-2}
\end{equation}
where the susceptible compartment is given by the direct integration of Eq.\ \ref{sir-generalgraph-massaction}
\begin{equation}
   S_k(t) = \left(1-h\right)\exp\left[-\mu_c k \phi_k(t) \right],
   \label{susceptibles-time-hmf}
\end{equation}
with initial conditions written in Eq. \ref{initial-conditions-hmf}. An analogous procedure also gives the following expression for the removed compartment
\begin{equation}
   R_k(t) = \mu_r \int_{0}^{t} I_k(t')dt'.
   \label{removed-time-hmf}
\end{equation}
In addition, we can write the auxiliary function given in Eq.\ \ref{auxiliary-function-2} by using Eq. \ref{removed-time-hmf} as
\begin{equation}
    \phi(t) = \frac{1}{\mu_{r}\langle k \rangle} \sum_{k} \left(\frac{k-1}{k}\right)P(k)R_{k}(t).
    \label{phi}
\end{equation}
Finally, by the differentiation of Eq. \ref{phi}, and by using Eqs.\ \ref{constraint-generalgraph}, and \ref{susceptibles-time-hmf}, we can write the following evolution equation
\begin{equation}
  \frac{\partial \phi}{\partial t} = \frac{1}{\left< k \right>}\left< \frac{k-1}{k} \right>  - \mu_r \phi(t)
        - \frac{1-h}{\left< k \right>} \sum_{k} \left( \frac{k-1}{k} \right)P(k)\exp\left[-\mu_c k \phi(t)\right].
  \label{sir-evolution-hmf}                         
\end{equation}

First, we discuss the asymptotic solution of Eq.\ \ref{sir-evolution-hmf} for $h=0$ and $\phi \to 0$. In this regime, we can expand the exponential in the last term of Eq.\ \ref{sir-evolution-hmf} up to first order to write
\begin{equation}
  \frac{\partial \phi}{\partial t} \approx \mu_r\left(\frac{\lambda}{\lambda_c} - 1\right) \phi(t)
  \label{sir-evolution-lowprevalence}                         
\end{equation}
where $\lambda_c$ is the critical basic reproduction number for the annealed networks, given by
\begin{equation}
  \lambda_c = \frac{\left< k \right>}{\left< k \right> - 1}.
  \label{critical-threshold}
\end{equation}
The solution for low prevalences are given by
\begin{equation}
  \phi(t) \approx \exp\left(t/\tau\right),
\end{equation}
where
\begin{equation}
  \tau = \frac{1}{\mu_r\left(\frac{\lambda}{\lambda_c} - 1 \right)},
\end{equation}
and for $\tau>0$, or $\lambda>\lambda_c$, we have a global epidemic state with exponential growth for low prevalences. Note that $\lambda_c$ separates the endemic and epidemic phases.

Now, we discuss the effects of a power-law network with distribution
\begin{equation}
P(k) = \begin{cases}
          \frac{\gamma-1}{f(\gamma-1)}m^{\gamma-1}k^{-\gamma},  & \quad \text{if } m \leq k \leq k_c; \\
          0,                                                    & \quad \text{if } k < m \text{ and if } k > k_c; \\
     \end{cases}
  \label{degree-distribution}
\end{equation}
where $f(x)$ is given by
\begin{equation}
  f(x) = 1 - \left( \frac{m}{k_c} \right)^x,
  \label{correction-f}
\end{equation}
which expresses the effect of a cutoff $k_c$. A scale-free network should have a cutoff in the degree distribution to become neutral, given by
\begin{equation}
  k_c = N^{1/\omega},
  \label{cutoff}
\end{equation}
where $\omega=2$ in the case of structural cutoff imposed on UCM networks\cite{Catanzaro-2005}, and $\omega = \gamma-1$ in the case of natural cutoff of BA networks, for example\cite{Barabasi-2016}. In addition, we can obtain the degree moments from the distribution in Eq. \ref{degree-distribution}
\begin{equation}
  \langle k^{\ell} \rangle = \frac{(\gamma -1)}{(\gamma-\ell-1)}\frac{f(\gamma-\ell-1)}{f(\gamma-1)}m^{\ell},
  \label{degree-moments}
\end{equation}
where $f(x)$, and $k_c$ are given by Eqs. \ref{correction-f}, and \ref{cutoff}, respectively.

We can obtain an approximation for the stationary solution. Taking the continuous limit of Eq.\ \ref{sir-evolution-hmf}, we can write the stationary solution as
\begin{eqnarray}
\mu_{c}\phi_{\infty} &=& 1 - \frac{(\gamma-1)}{\gamma}\frac{f(\gamma)}{f(\gamma-1)}\frac{1}{m} - \nonumber \\
   &-& \left(1-h\right)\frac{\gamma-2}{f(\gamma-2)}\frac{1}{m} \left( \mu_{c} m \phi_{\infty}\right)^{\gamma-1}
   \Gamma\left[-\gamma+1,\mu_{c}m\phi_{\infty}\right] + \nonumber \\
   &+& \left(1-h\right)\frac{\gamma-2}{f(\gamma-2)}\frac{1}{m^2} \left( \mu_{c} m \phi_{\infty}\right)^{\gamma}
   \Gamma\left[-\gamma,\mu_{c}m\phi_{\infty}\right] + \nonumber \\
   &+& \left(1-h\right)\left( \frac{m}{k_c} \right)^{\gamma-2}\frac{\gamma-2}{f(\gamma-2)}\frac{1}{k_c} 
       \left( \mu_{c} k_c \phi_{\infty}\right)^{\gamma-1}\Gamma\left[-\gamma+1,\mu_{c}k_c\phi_{\infty}\right] - \nonumber \\
   &-& \left(1-h\right)\left( \frac{m}{k_c} \right)^{\gamma-2}\frac{\gamma-2}{f(\gamma-2)}\frac{1}{k_c^2} 
       \left( \mu_{c} k_c \phi_{\infty}\right)^{\gamma}\Gamma\left[-\gamma,\mu_{c}k_c\phi_{\infty}\right].
   \label{stationary-solution-hmf}
\end{eqnarray}
where $\Gamma(\ell,x)$ is an incomplete Gamma function and $\phi_\infty$ is the asymptotic limit of $\phi(t)$ for $t\to\infty$. From the following asymptotic expansion of the incomplete Gamma function for $x \to 0$
\begin{equation}
  x^{\ell}\Gamma\left(-\ell,x\right) = \frac{\pi}{\sin\left[\pi(\ell+1)\right]} \frac{x^\ell}{\Gamma(\ell+1)}
  + \frac{1}{\ell} + \frac{x}{1-\ell} + \frac{x^2}{2\left(\ell-2\right)} + \cdots
  \label{gamma-expansion}
\end{equation}
we can write, up to second order in $\phi_\infty$
\begin{equation}
  h \approx \mu_r \left( 1 - \frac{\lambda}{\lambda_c}\right) \left< k \right> \phi_\infty 
  + \frac{\mu_c^2}{2}g\left< k \right>^2\phi_\infty^2.
  \label{scaling-phi}
\end{equation}
where we used the asymptotic behavior
\begin{equation}
  \left< \frac{k-1}{k}\right> \approx 1,
\end{equation}
valid for increasing $k$, and the scaling correction factor $g$ is
\begin{equation}
  g = \frac{\left< k^2 \right>}{\left<k\right>^2} - \frac{1}{\left<k\right>} \approx \frac{\left< k^2 \right>}{\left<k\right>^2},
  \label{correction-factor}
\end{equation}
which is the same scaling correction factor for the contact process\cite{Harris-1974} with temporary immunity\cite{Boguna-2009, Ferreira-2011a, Ferreira-2011}. In addition, $\lambda$, and $\lambda_c$ are given in Eqs.\ \ref{R0}, and \ref{critical-threshold}, respectively. The explicit form of $g$ is
\begin{equation}
  g = \frac{(\gamma-2)^2}{(\gamma-1)(\gamma-3)}\frac{f(\gamma-1)f(\gamma-3)}{f(\gamma-2)^2}.
  \label{correction-factor-explicit}
\end{equation}

We can obtain the stationary number of removed individuals from Eq.\ \ref{scaling-phi}. Using the fact that the system always evolve to an absorbing phase with $I_\infty = 0$, we can write from Eqs.\ \ref{constraint-generalgraph}, and \ref{susceptibles-time-hmf}
\begin{equation}
  R_\infty = 1 - (1-h)\sum_{k} P(k)\exp\left( \mu_{c}k \phi_{\infty}\right),
  \label{stationary-removed-hcm}
\end{equation}
and for a uncorrelated network with a degree distribution written in Eq. \ref{degree-distribution} in the continuous limit, we can write
\begin{eqnarray}
  R_\infty &=& 1 - \left(1-h\right)\frac{\gamma-1}{f(\gamma-1)} \left( \mu_{c} m \phi_{\infty}\right)^{\gamma-1}
       \Gamma\left[-\gamma+1,\mu_{c}m\phi_{\infty}\right] + \nonumber \\
  &+& \left(1-h\right)\left( \frac{m}{k_c} \right)^{\gamma-1}\frac{\gamma-1}{f(\gamma-1)} 
       \left( \mu_{c} k_c \phi_{\infty}\right)^{\gamma-1}\Gamma\left[-\gamma+1,\mu_{c}k_c\phi_{\infty}\right]
\end{eqnarray}
and again, from the asymptotic expansion in Eq. \ref{gamma-expansion}, we can write up to second order in $\phi_\infty$
\begin{equation}
  R_\infty - h \approx \mu_c \left< k \right> \phi_\infty - \frac{\mu_c^2}{2} \left< k^2 \right> \phi_\infty^2
\end{equation}
which we can invert to write, up to first order
\begin{equation}
  \phi_\infty \approx \frac{1}{\mu_c\left< k \right>}\left(R_\infty - h\right)
  \label{phi2}
\end{equation}
indicating that $\phi_\infty$ and $R_\infty$ possess the same scaling in the thermodynamic limit. Combining Eqs.\ \ref{scaling-phi}, and \ref{phi2}, we obtain
\begin{equation}
  h \approx \left( \frac{1}{\lambda} - \frac{1}{\lambda_c}\right) \left( R_\infty - h \right)
  + \frac{1}{2}g \left( R_\infty - h \right)^2.
  \label{scaling-removed}
\end{equation}
where $\lambda$, $\lambda_c$, and $g$ are given in Eqs.\ \ref{R0}, \ref{critical-threshold}, and \ref{correction-factor}, respectively. 

A similar process on obtaining Eqs. \ref{removed-scaling-wellstirring}, and \ref{shifting-scaling-wellstirring} yields
\begin{equation}
  R_\infty - \frac{1}{N} \approx \left( \frac{gN}{2} \right)^{1/2},
   \label{removed-scaling-scalefree}
\end{equation}
and
\begin{equation}
  \frac{1}{\lambda_c} - \frac{1}{\lambda} \approx \left( \frac{2N}{g} \right)^{1/2}.
  \label{shifting-scaling-scalefree}
\end{equation}
which we can summarize in the following scaling relation
\begin{equation}
  R_\infty - \frac{1}{N} \approx \left(\frac{gN}{2}\right)^{-1/2}F\left[\left(\frac{2N}{g}\right)^{1/2}\left(\frac{1}{\lambda_c} - \frac{1}{\lambda}\right) \right].
\label{scalefree-finitesize-scaling}
\end{equation}
and, finally, from the explicit expression for the correction factor $g$ in Eq.\ \ref{correction-factor-explicit}, we can write the mesoscopic dependence of $g$
\begin{equation}
  g = \begin{cases}
          \frac{N^{1/\omega}f(1)^2}{m\ln^2\left(\frac{m}{N^{1/\omega}}\right)},      & \quad \text{if } \gamma = 2; \\
          \frac{\left(\gamma-2\right)^2}{\left(\gamma-1\right)\left(3-\gamma\right)}
          \frac{f\left(\gamma-1\right)f\left(3-\gamma\right)}{f\left(\gamma-2\right)^2}
          \frac{N^{\frac{3-\gamma}{\omega}}}{m^{3-\gamma}},                          & \quad \text{if } 2 < \gamma < 3; \\
          -\frac{f(2)}{2f(1)^2}\ln\left(\frac{m}{N^{1/\omega}}\right)                & \quad \text{if } \gamma = 3; \\
          \frac{\left(\gamma-2\right)^2}{\left(\gamma-1\right)\left(\gamma-3\right)} 
          \frac{f\left(\gamma-1\right)f\left(\gamma-3\right)}{f\left(\gamma-2\right)^2} & \quad \text{if } \gamma > 3.
     \end{cases}
  \label{correction-factor-mesoscopic}
\end{equation}
It is interesting to note the presence of logarithm corrections in marginal values $\gamma=2$ and $\gamma=3$.

\section{Model and Results\label{sec:results}}

\begin{figure}[h!]
\begin{center}
\includegraphics[scale=0.15]{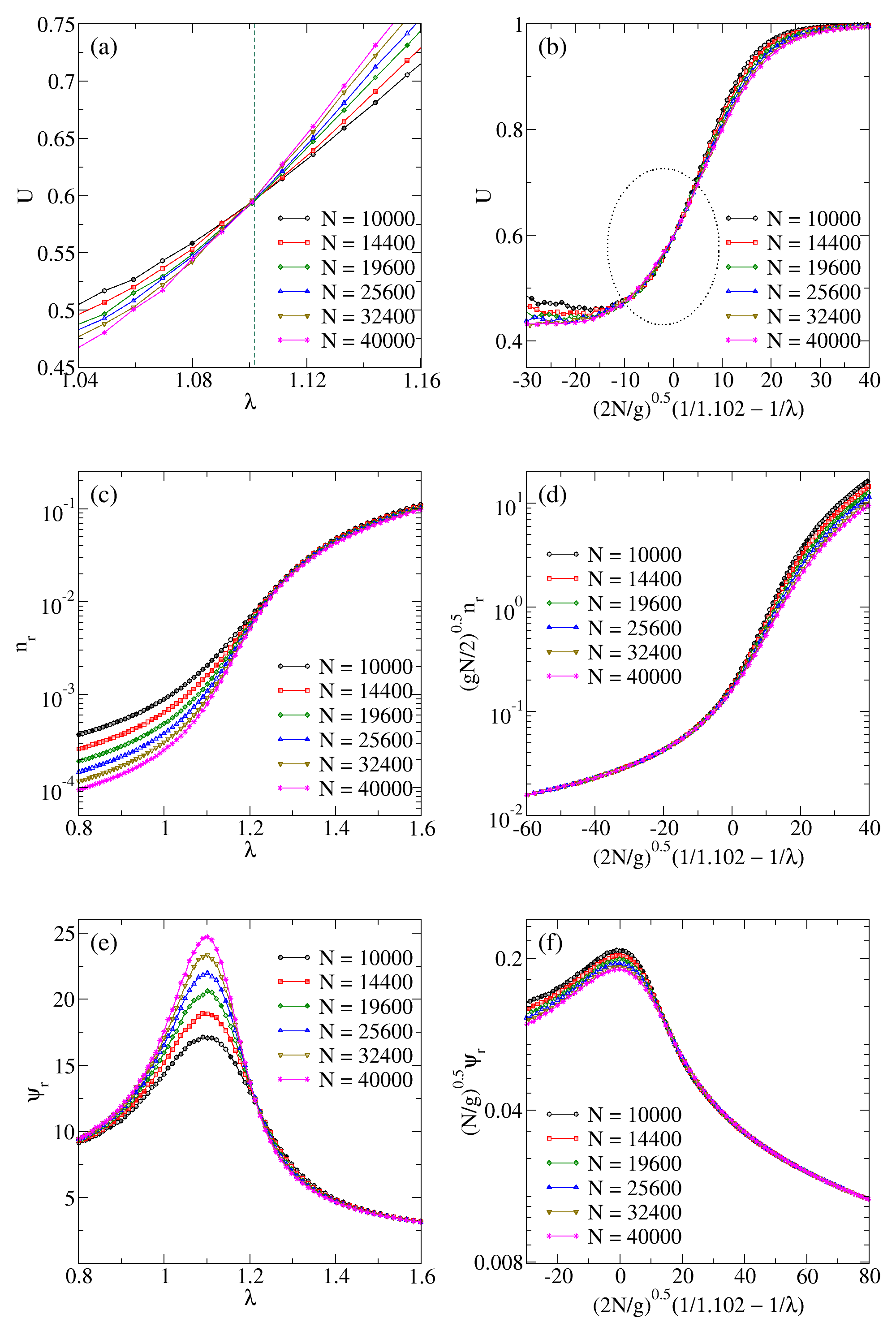}
\end{center}
\caption{(Color Online) We show the stationary simulation results of averages in Eq.\ \ref{averages-ucm}, in (a), (c), and (e). respectively, for the Asynchronous SIR model coupled to an UCM network with $\gamma=2.5$ and $m=8$, as functions of the control parameter $\lambda$. We show data collapses of $U$, $n_r$, and $\psi_r$, in (b), (d), and (f), respectively. The critical threshold is estimated as $\lambda_c=1.102$.}
\label{sir-ucm-results}
\end{figure}

We applied the same method to determine the critical threshold and exponents of the Asynchronous SIR model coupled to UCM networks with minimal degree $m$\cite{Catanzaro-2005}. To obtain numerical averages, we generated 160 random realizations of UCM networks. For each network, we grew $10^{5}$ clusters using the Asynchronous SIR dynamics, starting from one seed and stopping the growing process at the absorbing state, and calculated the averages
\begin{eqnarray}
  U      &=& \left[ \frac{\left<N_r^2\right>^2}{\left<N_r\right>\left<N_r^3\right>} \right], \nonumber \\
  n_r    &=& \left[ \left< N_r \right> \right]/N, \nonumber \\
  \psi_r &=& \left[ \frac{\left<N_r^2\right>}{\left<N_r\right>^2} \right],
  \label{averages-ucm}
\end{eqnarray}
where $\left[\cdots\right]$ means a quench average, and now $N_r$ is the number of removed individuals, excluding the seed. Following the results of the previous section, we propose the following finite-size scaling relations
\begin{eqnarray}
   U      &\approx& f_{U}\left[\left(2N/g\right)^{1/2}\left(1/\lambda_c-1/\lambda\right)\right], \nonumber \\
   n_r    &\approx& (gN/2)^{-1/2}f_{n_r}\left[\left(2N/g\right)^{1/2}\left(1/\lambda_c-1/\lambda\right)\right], \nonumber \\
   \psi_r &\approx& (2N/g)^{1/2}f_{\psi_r}\left[\left(2N/g\right)^{1/2}\left(1/\lambda_c-1/\lambda\right)\right].
   \label{fss-relations-scalefree}
\end{eqnarray}
We show results of Asynchronous SIR model coupled to UCM networks in Fig.\ \ref{sir-ucm-results}. All data collapses are compatible with the finite-size scaling theory presented in the previous section. In addition, we obtain the infinite mass cluster exponent $\beta$ by conjecturing the finite-size scaling of $\psi_r$. We obtained similar data collapses of Fig.\ \ref{sir-ucm-results} for another values of $\gamma$ in the interval $(2.5,3.5)$ (not shown here), which are also consistent with finite-size scaling relations in Eq.\ $\ref{fss-relations-scalefree}$.

\begin{figure}[h!]
\begin{center}
\includegraphics[scale=0.15]{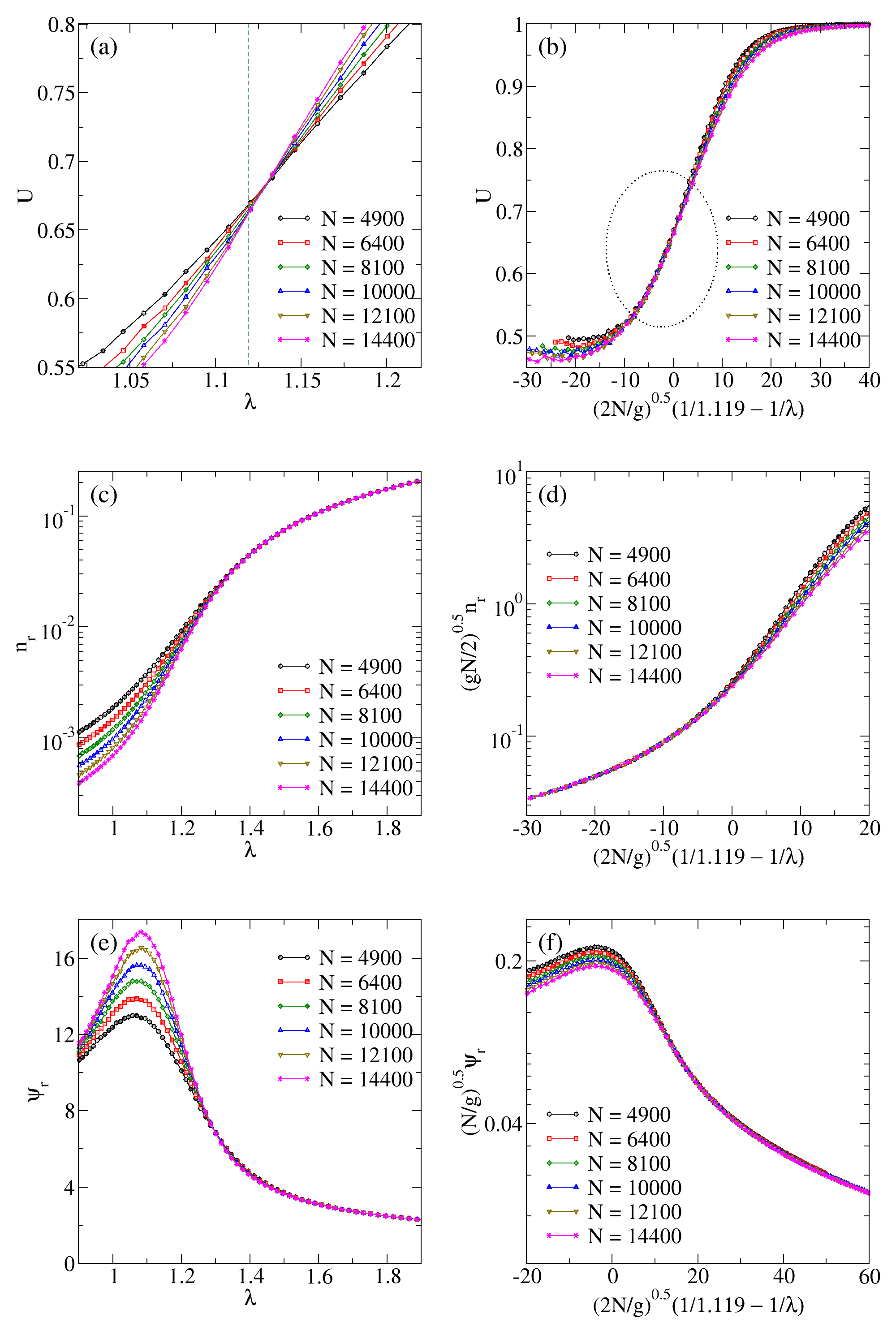}
\end{center}
\caption{(Color Online) We show the stationary simulation results of averages in Eq.\ \ref{averages-ucm}, in (a), (c), and (e). respectively, for the Asynchronous SIR model coupled to BA networks as functions of the control parameter $\lambda$. We show data collapses of $U$, $n_r$, and $\psi_r$, in (b), (d), and (f), respectively. The critical threshold is estimated as $\lambda_c=1.119$.}
\label{sir-barabasialbert}
\end{figure}

We also tested our finite-size scaling theory with our simulation results for a BA network with connectivity $z=10$ in Fig. \ref{sir-barabasialbert}, where the averages are obtained in an analogous way of UCM networks. Our data collapses are consistent with the mean-field regime with logarithmic corrections as written in Eq. \ref{fss-relations-scalefree} with $\gamma=3$ and $m=z/2$. We used the conjectured critical behavior to estimate the critical threshold by inspecting the best data collapses from the first and second moments of the removed density and the cumulant ratio. We repeated the same procedure for some network connectivities to collect the respective critical thresholds.

\begin{figure}[h!]
\begin{center}
\includegraphics[scale=0.25]{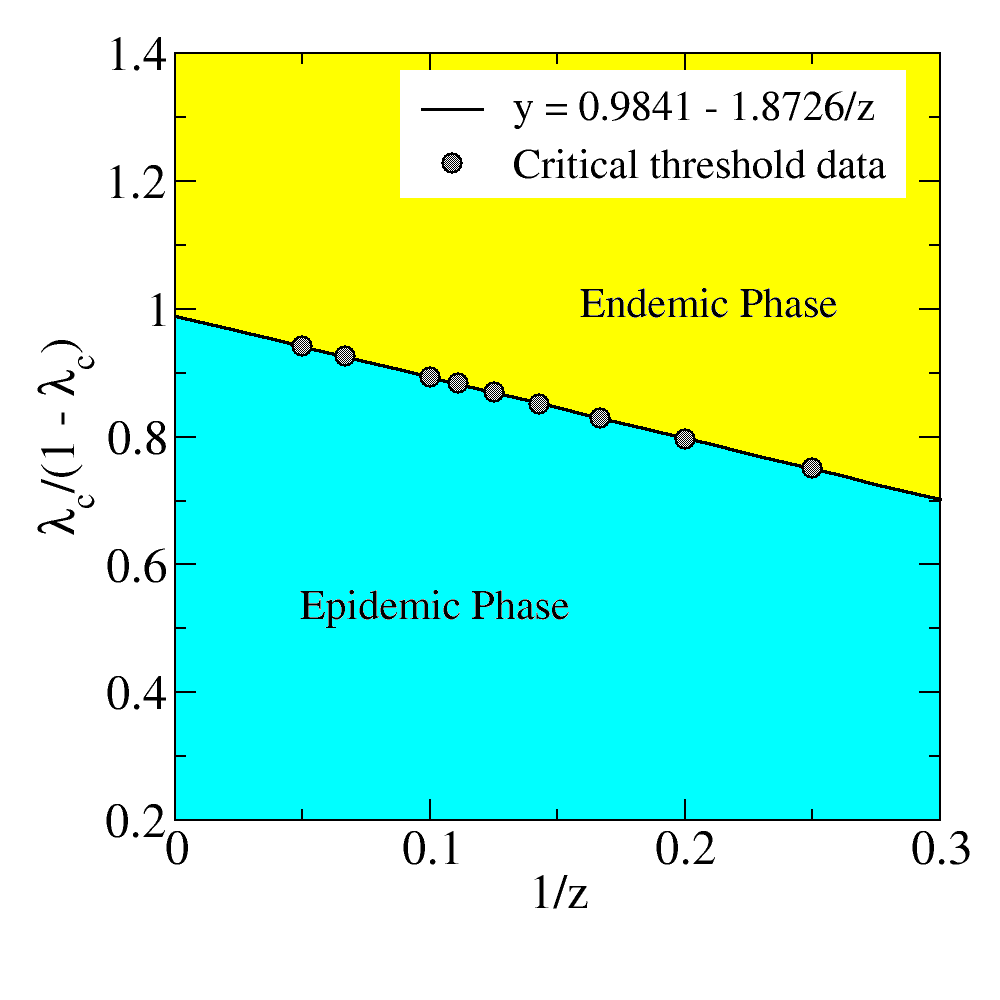}
\end{center}
\caption{(Color Online) We show the inverse of the critical basic reproduction number, i.e., $1/\lambda_c$ as a function of $1/z$ for the asynchronous SIR model on BA networks. Data presents a linear behavior. We can extrapolate data to obtain the well-stirring limit for $z \to \infty$ where the critical basic reproduction number is $\lambda_c=1$.}
\label{phase-diagram}
\end{figure}

In addition, we show the phase diagram of the model as a function of the network connectivity in Fig. \ref{phase-diagram}. We obtained a linear behavior of the critical $1/\lambda_c$ as a function of $1/z$. This behavior is consistent with Eq.\ \ref{critical-threshold} where we see that $1/\lambda_c$ is a linear function of $1/\left<k\right>$. We found the same behavior for other stochastic processes on the BA network\cite{Alves-2020, Alencar-2201.08708}, presenting mean-field critical behavior. Also particularly insightful is that a mean-field pair approximation yields the linear $1/z$ dependence of the critical $\lambda_c$ as seen, for example, for the contact process\cite{Tome-2015}. In the $z \to \infty$ limit, the model presents the expected critical basic reproduction number $\lambda_c=1$ of the complete graph\cite{Tome-2015} presented in the previous section.

We also obtained a mean-field critical behavior from the data collapses with logarithm corrections as predicted by the theory presented in the previous section. We can write from Eq. \ref{correction-factor} the asymptotic behavior of $g$ in the thermodynamic limit
\begin{equation}
g \propto \begin{cases}
          N^{1/\omega}\ln(N)^{-2},  & \quad \text{if } \gamma=2;   \\
          N^{(3-\gamma)/\omega},    & \quad \text{if } 2<\gamma<3; \\
          \ln(N),                   & \quad \text{if } \gamma=3;   \\
          \text{constant}           & \quad \text{if } \gamma>3.
          \end{cases}
  \label{correction-factor-scaling}
\end{equation}
In addition, when comparing the FSS relations in Eq.\ \ref{fss-relations-scalefree} with the FSS relations in Eq.\ \ref{fss-relations}, we can write for the effective critical exponent ratios for the first moment of the cluster distribution
\begin{equation}
  1 - \frac{\gamma'}{\nu} = \begin{cases}
          \frac{1}{2}+\frac{3-\gamma}{2\omega},  & \quad \text{if } 2 \leq \gamma <3;   \\
          \frac{1}{2},                           & \quad \text{if } \gamma \geq 3, 
          \end{cases}
  \label{removed-exponent}
\end{equation}
from the asymptotic behavior of the correction scaling $g$. From the conjectured scaling of the moment ratio $\psi_r$, we also have
\begin{equation}
  \frac{\beta}{\nu} = \begin{cases}
          \frac{1}{2}-\frac{\left(3-\gamma\right)}{2\omega},  & \quad \text{if } 2 \leq \gamma <3;   \\
          \frac{1}{2},                                        & \quad \text{if } \gamma \geq 3; 
          \end{cases}
  \label{2ndmoment-exponent}
\end{equation}
and from the shifting scaling, we obtain
\begin{equation}
  \frac{1}{\nu} = \begin{cases}
          \frac{1}{2}-\frac{3-\gamma}{2\omega},  & \quad \text{if } 2 \leq \gamma <3;   \\
          \frac{1}{2},                           & \quad \text{if } \gamma \geq 3.
          \end{cases}
  \label{shifting-exponent}
\end{equation}
Finally, by combining Eqs. \ref{removed-exponent}, \ref{2ndmoment-exponent}, and \ref{shifting-exponent}, we obtain $\beta=\gamma'=1$ for $2 \leq \gamma \leq 3$. The same result is valid for $\gamma \geq 3$. The $\gamma=2$ and $\gamma=3$ cases have additional logarithmic corrections.

\section{Conclusions\label{sec:conclusions}}

We presented the critical behavior of the Asynchronous SIR model on the UCM and BA scale-free networks. The model itself is essential in describing epidemic outbreaks like the COVID-19 pandemics. We defined the epidemic spreading by pairwise interactions where an infected individual can attack only one susceptible individual in a time interval, which differs from the definition of the standard SIR model \cite{Satorras-2001, Moreno-2002}, which allows one infected individual to attack all of his neighbors, the stochastic process defined in the present work provides a finite threshold. We also presented a finite-size scaling theory for the asynchronous SIR model on scale-free networks, where the finite-size dependence enters via the external field, defined as the initial infective density. An external field scaling as $1/N$ is interpreted as an random initial seed in the growing stochastic process.

In addition, we presented the dependence of the critical basic reproduction number $\lambda_c$ of the model on the network connectivity $z$ for BA networks. Our data is consistent with the Mean-Field dynamic percolation universality class exponents with strong logarithmic corrections. The $1/\lambda_c$ presents a linear dependence on $1/z$, which is consistent with the predicted behavior of the pair mean-field approximation seen in the results for the contact process on a generic graph with connectivity $z$. BA networks possess weak correlations, in a way they still obey the finite-size scaling presented in this manuscript for uncorrelated networks. By decreasing the network connectivity, we can increase the critical reproduction number, which is consistent with social distancing as a strategy to slow the epidemic spreading. Our results are also compatible with the $z \to \infty$ limit when we obtained by extrapolation the critical threshold of the model in the well-stirring limit.

\bibliography{textv1}

\begin{thebibliography}{25}%
\makeatletter
\providecommand \@ifxundefined [1]{%
 \@ifx{#1\undefined}
}%
\providecommand \@ifnum [1]{%
 \ifnum #1\expandafter \@firstoftwo
 \else \expandafter \@secondoftwo
 \fi
}%
\providecommand \@ifx [1]{%
 \ifx #1\expandafter \@firstoftwo
 \else \expandafter \@secondoftwo
 \fi
}%
\providecommand \natexlab [1]{#1}%
\providecommand \enquote  [1]{``#1''}%
\providecommand \bibnamefont  [1]{#1}%
\providecommand \bibfnamefont [1]{#1}%
\providecommand \citenamefont [1]{#1}%
\providecommand \href@noop [0]{\@secondoftwo}%
\providecommand \href [0]{\begingroup \@sanitize@url \@href}%
\providecommand \@href[1]{\@@startlink{#1}\@@href}%
\providecommand \@@href[1]{\endgroup#1\@@endlink}%
\providecommand \@sanitize@url [0]{\catcode `\\12\catcode `\$12\catcode
  `\&12\catcode `\#12\catcode `\^12\catcode `\_12\catcode `\%12\relax}%
\providecommand \@@startlink[1]{}%
\providecommand \@@endlink[0]{}%
\providecommand \url  [0]{\begingroup\@sanitize@url \@url }%
\providecommand \@url [1]{\endgroup\@href {#1}{\urlprefix }}%
\providecommand \urlprefix  [0]{URL }%
\providecommand \Eprint [0]{\href }%
\providecommand \doibase [0]{http://dx.doi.org/}%
\providecommand \selectlanguage [0]{\@gobble}%
\providecommand \bibinfo  [0]{\@secondoftwo}%
\providecommand \bibfield  [0]{\@secondoftwo}%
\providecommand \translation [1]{[#1]}%
\providecommand \BibitemOpen [0]{}%
\providecommand \bibitemStop [0]{}%
\providecommand \bibitemNoStop [0]{.\EOS\space}%
\providecommand \EOS [0]{\spacefactor3000\relax}%
\providecommand \BibitemShut  [1]{\csname bibitem#1\endcsname}%
\let\auto@bib@innerbib\@empty
%</preamble>
\bibitem [{\citenamefont {Cohen}\ and\ \citenamefont
  {Havlin}(2010)}]{Cohen-2010}%
  \BibitemOpen
  \bibfield  {author} {\bibinfo {author} {\bibfnamefont {R.}~\bibnamefont
  {Cohen}}\ and\ \bibinfo {author} {\bibfnamefont {S.}~\bibnamefont {Havlin}},\
  }\href@noop {} {\emph {\bibinfo {title} {Complex Networks: Structure,
  Robustness and Function}}}\ (\bibinfo  {publisher} {Cambridge University
  Press},\ \bibinfo {address} {Cambridge},\ \bibinfo {year} {2010})\BibitemShut
  {NoStop}%
\bibitem [{\citenamefont {Barab\'{a}si}\ and\ \citenamefont
  {P\'{o}sfai}(2016)}]{Barabasi-2016}%
  \BibitemOpen
  \bibfield  {author} {\bibinfo {author} {\bibfnamefont {A.-L.}\ \bibnamefont
  {Barab\'{a}si}}\ and\ \bibinfo {author} {\bibfnamefont {M.}~\bibnamefont
  {P\'{o}sfai}},\ }\href@noop {} {\emph {\bibinfo {title} {Network science}}}\
  (\bibinfo  {publisher} {Cambridge University Press},\ \bibinfo {address}
  {Cambridge},\ \bibinfo {year} {2016})\BibitemShut {NoStop}%
\bibitem [{\citenamefont {Keeling}\ and\ \citenamefont
  {Rohani}(2007)}]{Keeling-2007}%
  \BibitemOpen
  \bibfield  {author} {\bibinfo {author} {\bibfnamefont {M.}~\bibnamefont
  {Keeling}}\ and\ \bibinfo {author} {\bibfnamefont {P.}~\bibnamefont
  {Rohani}},\ }\href@noop {} {\emph {\bibinfo {title} {Modeling Infectious
  Diseases in Humans and Animals}}}\ (\bibinfo  {publisher} {Princeton
  University Press},\ \bibinfo {address} {Princeton},\ \bibinfo {year}
  {2007})\BibitemShut {NoStop}%
\bibitem [{\citenamefont {Keeling}\ and\ \citenamefont
  {Rohani}(2008)}]{Keeling-2008}%
  \BibitemOpen
  \bibfield  {author} {\bibinfo {author} {\bibfnamefont {M.}~\bibnamefont
  {Keeling}}\ and\ \bibinfo {author} {\bibfnamefont {P.}~\bibnamefont
  {Rohani}},\ }\href@noop {} {\emph {\bibinfo {title} {Modeling Infectious
  Diseases in Humans and Animals}}}\ (\bibinfo  {publisher} {Princeton
  University Press},\ \bibinfo {address} {Princeton},\ \bibinfo {year}
  {2008})\BibitemShut {NoStop}%
\bibitem [{\citenamefont {van Kampen}(1981)}]{vanKampen-1981}%
  \BibitemOpen
  \bibfield  {author} {\bibinfo {author} {\bibfnamefont {N.~G.}\ \bibnamefont
  {van Kampen}},\ }\href@noop {} {\emph {\bibinfo {title} {Stochastic Processes
  in Chemistry and Physics}}}\ (\bibinfo  {publisher} {North-Holland},\
  \bibinfo {address} {Amsterdam},\ \bibinfo {year} {1981})\BibitemShut
  {NoStop}%
\bibitem [{\citenamefont {Pastor-Satorras}\ and\ \citenamefont
  {Vespignani}(2001)}]{Satorras-2001}%
  \BibitemOpen
  \bibfield  {author} {\bibinfo {author} {\bibfnamefont {R.}~\bibnamefont
  {Pastor-Satorras}}\ and\ \bibinfo {author} {\bibfnamefont {A.}~\bibnamefont
  {Vespignani}},\ }\href@noop {} {\bibfield  {journal} {\bibinfo  {journal}
  {Phys. Rev. Lett.}\ }\textbf {\bibinfo {volume} {86}},\ \bibinfo {pages}
  {3200} (\bibinfo {year} {2001})}\BibitemShut {NoStop}%
\bibitem [{\citenamefont {Moreno}\ \emph {et~al.}(2002)\citenamefont {Moreno},
  \citenamefont {Pastor-Satorras},\ and\ \citenamefont
  {Vespignani}}]{Moreno-2002}%
  \BibitemOpen
  \bibfield  {author} {\bibinfo {author} {\bibfnamefont {Y.}~\bibnamefont
  {Moreno}}, \bibinfo {author} {\bibfnamefont {R.}~\bibnamefont
  {Pastor-Satorras}}, \ and\ \bibinfo {author} {\bibfnamefont {A.}~\bibnamefont
  {Vespignani}},\ }\href@noop {} {\bibfield  {journal} {\bibinfo  {journal}
  {Eur. Phys. J. B}\ }\textbf {\bibinfo {volume} {26}},\ \bibinfo {pages} {521}
  (\bibinfo {year} {2002})}\BibitemShut {NoStop}%
\bibitem [{\citenamefont {Tom\'{e}}\ and\ \citenamefont
  {Ziff}(2010)}]{Tome-2010}%
  \BibitemOpen
  \bibfield  {author} {\bibinfo {author} {\bibfnamefont {T.}~\bibnamefont
  {Tom\'{e}}}\ and\ \bibinfo {author} {\bibfnamefont {R.~M.}\ \bibnamefont
  {Ziff}},\ }\href@noop {} {\bibfield  {journal} {\bibinfo  {journal} {Phys.
  Rev. E}\ }\textbf {\bibinfo {volume} {82}},\ \bibinfo {pages} {051921}
  (\bibinfo {year} {2010})}\BibitemShut {NoStop}%
\bibitem [{\citenamefont {de~Souza}\ \emph {et~al.}(2011)\citenamefont
  {de~Souza}, \citenamefont {Tom\'{e}},\ and\ \citenamefont
  {Ziff}}]{deSouza-2011}%
  \BibitemOpen
  \bibfield  {author} {\bibinfo {author} {\bibfnamefont {D.~R.}\ \bibnamefont
  {de~Souza}}, \bibinfo {author} {\bibfnamefont {T.}~\bibnamefont {Tom\'{e}}},
  \ and\ \bibinfo {author} {\bibfnamefont {R.~M.}\ \bibnamefont {Ziff}},\
  }\href@noop {} {\bibfield  {journal} {\bibinfo  {journal} {J. Stat. Mech.}\
  }\textbf {\bibinfo {volume} {2011}},\ \bibinfo {pages} {27202} (\bibinfo
  {year} {2011})}\BibitemShut {NoStop}%
\bibitem [{\citenamefont {Tom\'{e}}\ and\ \citenamefont
  {de~Oliveira}(2011)}]{Tome-2011}%
  \BibitemOpen
  \bibfield  {author} {\bibinfo {author} {\bibfnamefont {T.}~\bibnamefont
  {Tom\'{e}}}\ and\ \bibinfo {author} {\bibfnamefont {M.~J.}\ \bibnamefont
  {de~Oliveira}},\ }\href@noop {} {\bibfield  {journal} {\bibinfo  {journal}
  {J. Phys. A}\ }\textbf {\bibinfo {volume} {44}},\ \bibinfo {pages} {095005}
  (\bibinfo {year} {2011})}\BibitemShut {NoStop}%
\bibitem [{\citenamefont {Tom\'{e}}\ and\ \citenamefont
  {de~Oliveira}(2015)}]{Tome-2015}%
  \BibitemOpen
  \bibfield  {author} {\bibinfo {author} {\bibfnamefont {T.}~\bibnamefont
  {Tom\'{e}}}\ and\ \bibinfo {author} {\bibfnamefont {M.~J.}\ \bibnamefont
  {de~Oliveira}},\ }\href@noop {} {\emph {\bibinfo {title} {Stochastic Dynamics
  and Irreversibility}}}\ (\bibinfo  {publisher} {Springer},\ \bibinfo
  {address} {Berlin},\ \bibinfo {year} {2015})\BibitemShut {NoStop}%
\bibitem [{\citenamefont {Santos}\ \emph {et~al.}(2020)\citenamefont {Santos},
  \citenamefont {Alves}, \citenamefont {Alves},\ and\ \citenamefont
  {Macedo-Filho}}]{Santos-2020}%
  \BibitemOpen
  \bibfield  {author} {\bibinfo {author} {\bibfnamefont {G.~B.~M.}\
  \bibnamefont {Santos}}, \bibinfo {author} {\bibfnamefont {T.~F.~A.}\
  \bibnamefont {Alves}}, \bibinfo {author} {\bibfnamefont {G.~A.}\ \bibnamefont
  {Alves}}, \ and\ \bibinfo {author} {\bibfnamefont {A.}~\bibnamefont
  {Macedo-Filho}},\ }\href@noop {} {\bibfield  {journal} {\bibinfo  {journal}
  {Phys. Lett. A}\ }\textbf {\bibinfo {volume} {384}},\ \bibinfo {pages}
  {126063} (\bibinfo {year} {2020})}\BibitemShut {NoStop}%
\bibitem [{\citenamefont {Alencar}\ \emph {et~al.}(2020)\citenamefont
  {Alencar}, \citenamefont {Alves}, \citenamefont {Alves},\ and\ \citenamefont
  {Macedo-Filho}}]{Alencar-2020}%
  \BibitemOpen
  \bibfield  {author} {\bibinfo {author} {\bibfnamefont {D.~S.~M.}\
  \bibnamefont {Alencar}}, \bibinfo {author} {\bibfnamefont {T.~F.~A.}\
  \bibnamefont {Alves}}, \bibinfo {author} {\bibfnamefont {G.~A.}\ \bibnamefont
  {Alves}}, \ and\ \bibinfo {author} {\bibfnamefont {A.}~\bibnamefont
  {Macedo-Filho}},\ }\href@noop {} {\bibfield  {journal} {\bibinfo  {journal}
  {Physica A}\ }\textbf {\bibinfo {volume} {541}},\ \bibinfo {pages} {122800}
  (\bibinfo {year} {2020})}\BibitemShut {NoStop}%
\bibitem [{\citenamefont {Catanzaro}\ \emph {et~al.}(2005)\citenamefont
  {Catanzaro}, \citenamefont {Bogun\'{a}},\ and\ \citenamefont
  {Pastor-Satorras}}]{Catanzaro-2005}%
  \BibitemOpen
  \bibfield  {author} {\bibinfo {author} {\bibfnamefont {M.}~\bibnamefont
  {Catanzaro}}, \bibinfo {author} {\bibfnamefont {M.}~\bibnamefont
  {Bogun\'{a}}}, \ and\ \bibinfo {author} {\bibfnamefont {R.}~\bibnamefont
  {Pastor-Satorras}},\ }\href@noop {} {\bibfield  {journal} {\bibinfo
  {journal} {Phys. Rev. E}\ }\textbf {\bibinfo {volume} {71}},\ \bibinfo
  {pages} {027103} (\bibinfo {year} {2005})}\BibitemShut {NoStop}%
\bibitem [{\citenamefont {Ferguson}(2007)}]{Ferguson-2007}%
  \BibitemOpen
  \bibfield  {author} {\bibinfo {author} {\bibfnamefont {N.}~\bibnamefont
  {Ferguson}},\ }\href@noop {} {\bibfield  {journal} {\bibinfo  {journal}
  {Nature}\ }\textbf {\bibinfo {volume} {446}},\ \bibinfo {pages} {733}
  (\bibinfo {year} {2007})}\BibitemShut {NoStop}%
\bibitem [{\citenamefont {Bogun\'{a}}\ \emph {et~al.}(2013)\citenamefont
  {Bogun\'{a}}, \citenamefont {Castellano},\ and\ \citenamefont
  {Pastor-Satorras}}]{Boguna-2013}%
  \BibitemOpen
  \bibfield  {author} {\bibinfo {author} {\bibfnamefont {M.}~\bibnamefont
  {Bogun\'{a}}}, \bibinfo {author} {\bibfnamefont {C.}~\bibnamefont
  {Castellano}}, \ and\ \bibinfo {author} {\bibfnamefont {R.}~\bibnamefont
  {Pastor-Satorras}},\ }\href@noop {} {\bibfield  {journal} {\bibinfo
  {journal} {Phys. Rev. Lett.}\ }\textbf {\bibinfo {volume} {111}},\ \bibinfo
  {pages} {068701} (\bibinfo {year} {2013})}\BibitemShut {NoStop}%
\bibitem [{\citenamefont {Stauffer}\ and\ \citenamefont
  {Aharony}(1992)}]{Stauffer-1992}%
  \BibitemOpen
  \bibfield  {author} {\bibinfo {author} {\bibfnamefont {D.}~\bibnamefont
  {Stauffer}}\ and\ \bibinfo {author} {\bibfnamefont {A.}~\bibnamefont
  {Aharony}},\ }\href@noop {} {\emph {\bibinfo {title} {Introduction to
  Percolation Theory}}}\ (\bibinfo  {publisher} {Taylor \& Francis},\ \bibinfo
  {address} {London},\ \bibinfo {year} {1992})\BibitemShut {NoStop}%
\bibitem [{\citenamefont {Christensen}\ and\ \citenamefont
  {Moloney}(2005)}]{Christensen-2005}%
  \BibitemOpen
  \bibfield  {author} {\bibinfo {author} {\bibfnamefont {K.}~\bibnamefont
  {Christensen}}\ and\ \bibinfo {author} {\bibfnamefont {N.~R.}\ \bibnamefont
  {Moloney}},\ }\href@noop {} {\emph {\bibinfo {title} {Complexity and
  Criticality}}}\ (\bibinfo  {publisher} {Imperial College Press},\ \bibinfo
  {address} {London},\ \bibinfo {year} {2005})\BibitemShut {NoStop}%
\bibitem [{\citenamefont {Barrat}\ \emph {et~al.}(2008)\citenamefont {Barrat},
  \citenamefont {Barth{\'e}lemy},\ and\ \citenamefont
  {Vespignani}}]{Barrat-2008}%
  \BibitemOpen
  \bibfield  {author} {\bibinfo {author} {\bibfnamefont {A.}~\bibnamefont
  {Barrat}}, \bibinfo {author} {\bibfnamefont {M.}~\bibnamefont
  {Barth{\'e}lemy}}, \ and\ \bibinfo {author} {\bibfnamefont {A.}~\bibnamefont
  {Vespignani}},\ }\href@noop {} {\emph {\bibinfo {title} {Dynamical Processes
  on Complex Networks}}}\ (\bibinfo  {publisher} {Cambridge University Press},\
  \bibinfo {address} {Cambridge},\ \bibinfo {year} {2008})\BibitemShut
  {NoStop}%
\bibitem [{\citenamefont {Harris}(1974)}]{Harris-1974}%
  \BibitemOpen
  \bibfield  {author} {\bibinfo {author} {\bibfnamefont {T.~E.}\ \bibnamefont
  {Harris}},\ }\href@noop {} {\bibfield  {journal} {\bibinfo  {journal} {Ann.
  Prob.}\ }\textbf {\bibinfo {volume} {2}},\ \bibinfo {pages} {969} (\bibinfo
  {year} {1974})}\BibitemShut {NoStop}%
\bibitem [{\citenamefont {Bogun\'{a}}\ \emph {et~al.}(2009)\citenamefont
  {Bogun\'{a}}, \citenamefont {Castellano},\ and\ \citenamefont
  {Pastor-Satorras}}]{Boguna-2009}%
  \BibitemOpen
  \bibfield  {author} {\bibinfo {author} {\bibfnamefont {M.}~\bibnamefont
  {Bogun\'{a}}}, \bibinfo {author} {\bibfnamefont {C.}~\bibnamefont
  {Castellano}}, \ and\ \bibinfo {author} {\bibfnamefont {R.}~\bibnamefont
  {Pastor-Satorras}},\ }\href@noop {} {\bibfield  {journal} {\bibinfo
  {journal} {Phys. Rev. E}\ }\textbf {\bibinfo {volume} {79}},\ \bibinfo
  {pages} {036110} (\bibinfo {year} {2009})}\BibitemShut {NoStop}%
\bibitem [{\citenamefont {Ferreira}\ \emph
  {et~al.}(2011{\natexlab{a}})\citenamefont {Ferreira}, \citenamefont
  {Ferreira},\ and\ \citenamefont {Pastor-Satorras}}]{Ferreira-2011a}%
  \BibitemOpen
  \bibfield  {author} {\bibinfo {author} {\bibfnamefont {S.~C.}\ \bibnamefont
  {Ferreira}}, \bibinfo {author} {\bibfnamefont {R.~S.}\ \bibnamefont
  {Ferreira}}, \ and\ \bibinfo {author} {\bibfnamefont {R.}~\bibnamefont
  {Pastor-Satorras}},\ }\href@noop {} {\bibfield  {journal} {\bibinfo
  {journal} {Phys. Rev. E}\ }\textbf {\bibinfo {volume} {83}},\ \bibinfo
  {pages} {066113} (\bibinfo {year} {2011}{\natexlab{a}})}\BibitemShut
  {NoStop}%
\bibitem [{\citenamefont {Ferreira}\ \emph
  {et~al.}(2011{\natexlab{b}})\citenamefont {Ferreira}, \citenamefont
  {Ferreira}, \citenamefont {Castellano},\ and\ \citenamefont
  {Pastor-Satorras}}]{Ferreira-2011}%
  \BibitemOpen
  \bibfield  {author} {\bibinfo {author} {\bibfnamefont {S.~C.}\ \bibnamefont
  {Ferreira}}, \bibinfo {author} {\bibfnamefont {R.~S.}\ \bibnamefont
  {Ferreira}}, \bibinfo {author} {\bibfnamefont {C.}~\bibnamefont
  {Castellano}}, \ and\ \bibinfo {author} {\bibfnamefont {R.}~\bibnamefont
  {Pastor-Satorras}},\ }\href@noop {} {\bibfield  {journal} {\bibinfo
  {journal} {Phys. Rev. E}\ }\textbf {\bibinfo {volume} {84}},\ \bibinfo
  {pages} {066102} (\bibinfo {year} {2011}{\natexlab{b}})}\BibitemShut
  {NoStop}%
\bibitem [{\citenamefont {Alves}\ \emph {et~al.}(2020)\citenamefont {Alves},
  \citenamefont {Alves}, \citenamefont {Lima},\ and\ \citenamefont
  {Macedo-Filho}}]{Alves-2020}%
  \BibitemOpen
  \bibfield  {author} {\bibinfo {author} {\bibfnamefont {T.~F.~A.}\
  \bibnamefont {Alves}}, \bibinfo {author} {\bibfnamefont {G.~A.}\ \bibnamefont
  {Alves}}, \bibinfo {author} {\bibfnamefont {F.~W.~S.}\ \bibnamefont {Lima}},
  \ and\ \bibinfo {author} {\bibfnamefont {A.}~\bibnamefont {Macedo-Filho}},\
  }\href@noop {} {\bibfield  {journal} {\bibinfo  {journal} {J. Stat. Mech.}\
  }\textbf {\bibinfo {volume} {2020}},\ \bibinfo {pages} {033203} (\bibinfo
  {year} {2020})}\BibitemShut {NoStop}%
\bibitem [{\citenamefont {Alencar}\ \emph {et~al.}(2022)\citenamefont
  {Alencar}, \citenamefont {Alves}, \citenamefont {Alves}, \citenamefont
  {Ferreira}, \citenamefont {Macedo-Filho},\ and\ \citenamefont
  {Lima}}]{Alencar-2201.08708}%
  \BibitemOpen
  \bibfield  {author} {\bibinfo {author} {\bibfnamefont {D.~S.~M.}\
  \bibnamefont {Alencar}}, \bibinfo {author} {\bibfnamefont {T.~F.~A.}\
  \bibnamefont {Alves}}, \bibinfo {author} {\bibfnamefont {G.~A.}\ \bibnamefont
  {Alves}}, \bibinfo {author} {\bibfnamefont {R.~S.}\ \bibnamefont {Ferreira}},
  \bibinfo {author} {\bibfnamefont {A.}~\bibnamefont {Macedo-Filho}}, \ and\
  \bibinfo {author} {\bibfnamefont {F.~W.~S.}\ \bibnamefont {Lima}},\ }\href
  {\doibase 10.48550/ARXIV.2201.08708} {\enquote {\bibinfo {title} {Phase
  diagram of the contact process on barabasi-albert networks},}\ } (\bibinfo
  {year} {2022})\BibitemShut {NoStop}%
\end{thebibliography}%

\end{document}